\documentstyle[12pt,a4]{article}

\newcommand{\tnm}{{\cal T}_{nm}}
\newcommand{\bea}{\begin{eqnarray}}
\newcommand{\eea}{\end{eqnarray}}
\newcommand{\be}{\begin{equation}}
\newcommand{\ee}{\end{equation}}
\newcommand{\pkt}{\; .}
\newcommand{\kma}{\; ,}
\newcommand{\tdot}[1]{\stackrel{\dots}{#1}}

\newcommand{\re}{{\rm Re}}

\newcommand{\abo}{\partial^\mu}
\newcommand{\abu}{\partial_\mu}

\newcommand{\om}{\omega_{k0}}
\newcommand{\intk}{\int\!\frac{{d^3}k }{(2\pi)^32\om}\,}
\newcommand{\intko}{\int\!\frac{{d^3}k }{(2\pi)^3\,}}

\newcommand{\calm}{{\cal M}}
\newcommand{\delm}{\delta m}
\newcommand{\dellam}{\delta \lambda}
\newcommand{\calv}{{\cal V}}
\newcommand{\calff}{{\cal F}_{\rm fin}}
\newcommand{\calfft}{\tilde\calff}
\newcommand{\imone}{I_{-1}}
\newcommand{\imthree}{I_{-3}}
\newcommand{\cc}{{\cal C}}

\newcommand{\calf}{{\cal F}}

\newcommand{\calE}{{\cal E}}

\newcommand{\bfk}{{\bf k}}
\newcommand{\bfx}{{\bf x}}

\begin{document}
\begin{titlepage}
\begin{flushright}
DO-TH-97/29\\
December 1997
\end{flushright}
\vspace{20mm}
\begin{center}
{\Large \bf
Renormalization of nonequilibrium dynamics at large $N$ and
finite temperature }\\
\vspace{10mm}
{\large  J\"urgen Baacke\footnote{
 e-mail:~baacke@physik.uni-dortmund.de}, Katrin Heitmann
\footnote{e-mail:~heitmann@hal1.physik.uni-dortmund.de}, and
Carsten P\"atzold
\footnote{e-mail:~paetzold@hal1.physik.uni-dortmund.de}} \\
\vspace{15mm}

{\large Institut f\"ur Physik, Universit\"at Dortmund} \\
{\large D - 44221 Dortmund, Germany}
\vspace{25mm}

{\bf Abstract}
\end{center} 
We generalize a previously proposed renormalization and
computation scheme for nonequilibrium dynamics to
include finite temperature and one-loop selfconsistency
as arising in the large-$N$ limit. Since such a scheme
amounts essentially to tadpole resummation, it also includes,
at high temperature, the hard mass corrections proportional to
$T^2$. We present some numerical examples at $T=0$ and for
finite temperature; the results reproduce the essential features
of those of other groups. Especially we can confirm a recently 
discovered sum rule for the late time behaviour. 
\end{titlepage}


\section{Introduction}
Nonequilibrium dynamics in quantum field theory 
has become, during the last years, a
very active field of research in particle physics [1-5],
in cosmology [6-18]
, and in solid state physics \cite{Zurek:1996}.
The application of the general formalism of nonequilibrium
quantum field theory \cite{Schwinger:1961,Keldysh:1964}
has been limited up to now mostly to spatially homogeneous systems
(see however \cite{Lampert:1996}).
The typical field theoretic system considered in this context
consists in a classical,
spatially constant field $\phi(t)$ (Higgs, inflaton, condensate)
and a quantum state of fluctuations of the same or another field,
chosen initially as a Bogoliubov-transformed vacuum state
or a thermal state. The classical field is started
with an initial value away from
a local or global minimum of the classical or effective action.
The time development of this coupled system is then studied including
the back reaction of the quantum field in one-loop, Hartree or
large-$N$ approximations.
The basic equations of motions have been derived by several groups
and a considerable number of numerical studies has been performed.
It has been found that the system is far from being Markovian,
showing a long-time memory. The typical late-time
behaviour is a stationary oscillation of the field
$\phi(t)$ and of the quantum fluctuations \cite{Boyanovsky:1996,
deVega:1997,Cooper:1997}. 
Contrary to naive expections the classical field
does not come to rest and the quantum fluctuations do not
thermalize in the various approximations that have been studied.

The question of a concise and covariant renormalization may not 
be the most urgent one in some of the above-mentioned contexts.
Logarithmic corrections in the fluctuation integral, typically with
coefficients $\lambda/16\pi^2$, are of course small
in most cases; covariance may not be important in solid state
applications; and in the case of effective theories like
the sigma model for disordered chiral condensates
renormalization may become replaced by physical cutoffs.
Also, due to parametric resonance, suppressed by
the back reaction,  quantum
fluctuations develop most strongly in the small momentum 
region; so a cutoff chosen with taste will do it for
practical purposes. Nevertheless, in an expanding 
field of research as nonequilibrium
quantum field theory one should make sure that things can be done
properly. In a - still unexplored - GUT phase transition
the coupling will not be as small as for the inflationary
models being investigated at present.
Renormalization of the large-$N$ approximation 
certainly will be important as a basic step 
if one tries to include a real rescattering
of quantum fluctuations; such a rescattering - not only
through the back reaction with the classical field - 
is presumably an important ingredient for understanding
thermalization. A subject that arises in close connection
with renormalization are singularities in the time variable due to
initial conditions; it is a rather fundamental subject for
nonequilibrium dynamics. We have addressed this problem
recently \cite{Baacke:1997d}. Finally, 
from a practical point of view, 
our renormalization procedure also implies improving 
considerably the convergence of momentum integrals,
and the gain in computing time
 can be of importance when studying more complex
systems.

The basic method for the perturbative expansion
which we will use here has been developed in
\cite{Baacke:1997a}. It is based essentially on a 
standard resolvent 
expansion of the progagator. Most other groups have used
the eikonal expansion which seems a natural choice
in the presence of an 
oscillating background field and has its merit in allowing
the study of adiabatic properties \cite{Cooper:1997}. However,
renormalization becomes more cumbersome as the relation to
the Feynman graph expansion becomes more remote. Furthermore,
eikonal expansions cannot be generalized to coupled 
fields (one would need time-ordered exponentials, 
not suited for numerical computations).

We will present in the next section the $O(N)$ model, the 
nonequilibrium equations
of motion for the classical field and the fluctuation modes,
the energy density and the pressure in large-$N$
approximation and in unrenormalized form.
In order to prepare the discussion of renormalization we
give the basic equations for a perturbative expansion of
the mode equations in section 3. Large-N renormalization, including
finite temperature corrections is derived in section 4.
Section 5 is devoted to the high temperature 
limit of the model. Some numerical experiments are   
presented in section 6. A resum\'{e} and conclusions
are given in section 7.


\section{Formulation of the large-$N$ equations}
\setcounter{equation}{0}
We consider the $O(N)$ with  the
Lagrangian \footnote{We deviate from the usual convention
of introducing the interaction term with a factor $\lambda/8N$
in order to avoid a plethora of fractions $\frac{\lambda}{2}$ and,
even worse $\frac{\lambda+\dellam}{2}$,
 in the subsequent formulae.}
\be
\label{lagrange}
{\cal L}={\displaystyle{\frac 1 2}}\abu \phi^i\abo\phi^i
-{\displaystyle\frac 1 2}
m^2\phi^i\phi^i-\frac{\lambda}{4N}(\phi^i\phi^i)^2
\ee
where $\phi^i, i=1,..,N$ are $N$ real scalar fields.
The nonequilibrium state of the system is characterized
by a classical expectation value which we take in 
the direction of $\phi_N$. We split the field
into its expectation value $\phi$ and the quantum fluctuations
$\psi$ via
\begin{equation}
\label{erw}
\phi^i(\bfx,t)=\delta^i_N \sqrt{N} \phi(t)+\psi^i(\bfx,t)\pkt
\end{equation}
In the large $N$ limit one neglects, in the Lagrangian,
all terms which are not of order $N$. In particular
terms containing the fluctuation $\psi_N$ of
the component $\phi_N$ are at most of order $\sqrt{N}$ and are
dropped, therefore. This is in contrast to the Hartree approximation
where the fluctuations of $\phi_N$ are included. The fluctuations of
the other components are identical, their summation produces
factors $N-1=N(1+O(1/N))$.  Identifying all the fields
$\psi_1,..\psi_{N-1}$
as $\psi$ the leading order term in the Lagrangian then takes the form
\be \label{lnl1}
{\cal L}=N\left({\cal L}_{\phi}+{\cal L}_{\psi} + {\cal L}_I\right)\kma
\ee
with
\bea \label{lnl2}
{\cal L}_{\phi}&=&
\frac{1}{2}
\partial_\mu\phi\partial^\mu\phi-\frac{1}{2}m^2\phi^2-
\frac{\lambda}{4}\phi^4 \; ,\\
{\cal L}_{\psi}&=&\frac{1}{2}
\partial_\mu\psi\partial^\mu\psi-\frac{1}{2}m^2\psi^2
+\frac{\lambda}{4}(\psi^2)^2 \kma \\
{\cal L}_{\rm I}&=&
-\frac{\lambda}{2}\psi^2\phi^2 \; ,
\eea
where $\psi^2$ is to be identified with $\sum \psi^i \psi^i/N$.

We decompose the fluctuating field into momentum
eigenfunctions via
\be
\psi(\bfx,t)=\intk \left[ a_\bfk U_k(t) e^{i\bfk\bfx}
+a^\dagger_\bfk U^*_k(t) e^{-i\bfk\bfx}\right]\pkt
\ee
with $\om=\sqrt{m_0^2+k^2}$. The mass $m_0$ will be specified below.
This field decomposition defines a vacuum state as being annihiliated
by the operators $a_\bfk$.

The equations of motion for the field $\phi(t)$ 
and of the fluctuations $U_k(t)$ have been derived in this
formalism by
various authors \cite{Boyanovsky:1994,Cooper:1994,Cooper:1995}.
In addition to the large-$N$ Lagrangian, Eqs. (\ref{lnl1},
\ref{lnl2}),
one uses, on averaging over the quantum fluctuations, rules
like
\bea
(\psi^2)^2   &\Rightarrow& \langle \psi^2\rangle ^2 \kma\\
\frac{\partial(\psi^2)^2}{\partial\psi} &\Rightarrow
& 4 \psi \langle \psi^2\rangle
\kma \;\; {\rm or}
\\
\frac{\partial^2(\psi^2)^2}{\partial\psi^2} &\Rightarrow
&4\langle\psi^2\rangle
\eea
which follow at large $N$
from the identification $\psi^2\simeq\sum\psi^i\psi^i/N$.

We include in the following the counter terms that we will need
later in order to write the renormalized equations. So the
equation of motion for the field $\phi$ becomes
\be
\ddot{\phi}(t)+(m^2+\delta m^2) \phi(t)+
(\lambda+\dellam)\phi(t)
\left[\phi^2(t)+\calf (t,T)\right]=0\pkt
\ee
Here $\calf (t,T)$ is the divergent fluctuation integral; it is
given by the average of the fluctuation fields defined by the
initial density matrix. For a thermal initial state of quanta with
energy $\omega_{k0}=\sqrt{k^2+m_0^2}$ it is given by
\be
\calf(t,T)=\langle \psi^2 (\bfx,t)\rangle
=\intk \coth{\frac{\beta\om}{2}}{|U_k(t)|^2}\pkt
\ee
The mode functions satisfy the equation:
\be
\label{mode}
\left[\frac{d^2}{dt^2}+\omega_k^2(t)\right]U_k(t)=0\kma
\ee
and the initial conditions
\be
U_k(0)=1 \;\; ;\;\; \dot U_k (0)= - i \om \pkt
\ee
The time dependent frequency $ \omega_k (t)$ is given by
\be
\omega^2_k = k^2 +\calm^2(t)
\ee
with the time dependent mass
\be \label{effm}
\calm^2(t)=m^2+(\lambda+\dellam)
 \left[\phi^2(t)+\calf(t)\right]\pkt
\ee
As in our previous work we rewrite the mode equation in the form
\be
\left[\frac{d^2}{dt^2}+\omega_{k0}^2\right]
U_k(t)=-\calv(t)U_k(t)\kma
\ee
whereby we have defined the time-dependent potential
$\calv(t)={\cal M}^2(t)-{\cal M}^2(0)$; we further identify
$m_0=\calm (0)$ as the ``initial mass''.
The classical equation of motion also can be rewritten as
\be
\label{eqmotion}
\ddot{\phi}(t)+{\cal M}^2(t)\phi(t)=0\kma
\ee
which is of the same form as Eq. (\ref{mode}) with $k=0$, the
classical field also is referred to as ``zero mode'' in
\cite{Boyanovsky:1995}.
The average of energy with respect to the initial density matrix
is given by \footnote{Note that twice the last term, with positive sign,
is included in the fluctuation energy, since $\omega^2_k(t)$
contains $\calf(t,T)$.}
\bea
\label{energy}
{\cal E}&=&\frac{1}{2}\dot{\phi}^2(t)+{\displaystyle {\frac 1 2}}
(m^2+\delta m^2)\phi^2(t)
+\frac{\lambda+\dellam}{4}\phi^4(t)+\delta\Lambda
\nonumber\\
&&+\intk
\coth{\frac{\beta\om}{2}}
\left\{{\displaystyle \frac 1 2}|\dot{U}_k(t)|^2
+{\displaystyle \frac{1}{2}}\omega^2_k(t)|U_k(t)|^2\right\}
\\ \nonumber &&-
\frac{\lambda+\dellam}{4}\calf^2(t,T)\
\pkt
\eea
It is easy to check, using the equations of motion
(\ref{eqmotion}) and (\ref{mode}), that the energy is conserved.
The energy density is the $00$ component of the energy-momentum tensor.
The average of the energy momentum tensor for our system is diagonal,
its space-space components define the pressure which is given by
\bea \label{pdef}
p &=& \phi^2(t)-\calE + A \frac{d^2}{dt^2}
\left[\phi^2(t)+\calf(t,T)\right]\\ \nonumber
&&+\intk \coth{\frac{\beta\om}{2}} \left(\om^2+\frac{k^2}{3}\right)
|U_k(t)|^2 \pkt
\eea
The term proportional to $A$ is the space-space
component of the ``improvement'' term
$A(g_{\mu\nu}\partial^2-\partial_\mu\partial_\nu)\phi^2$ for the
energy momentum tensor as
introduced by \cite{Callan:1971}. It serves as a renormalization
counter term, here.


\section{Perturbative expansion}
\setcounter{equation}{0}
In order to prepare the renormalized version of the equations we 
introduce a suitable expansion of the mode functions. We have
used this method exhaustively in our previous publications for the
inflaton field coupled to itself \cite{Baacke:1997a} and to gauge
bosons \cite{Baacke:1997b} in Minkowski-space and
for the inflaton field coupled to itself in a conformally flat
FRW-universe \cite{Baacke:1997c}. 
All these calculations have been done for $T=0$.
The renormalization procedure does not change for $T\neq 0$. 
Therefore we give here only a brief review of the perturbative
expansion. For details the reader is referred to our previous work.

The mode functions can be written as
\begin{equation} 
\label{udgl}
\left[ \frac{d^2}{dt^2}+
\omega_{k0}^2\right]U_k(t)=-{\cal V}(t)U_k(t)\; ,
\end{equation}
with
\begin{eqnarray}
\label{potential}
{\cal V}(t)&=&{\cal M}^2(t)-{\cal M}^2(0)\; ,
\\
\omega_{k0}&=&\left[\vec k^2+{\cal M}^2(0)\right]^{1/2}
\end{eqnarray}
(for the definition of ${\cal M}^2(t)$ see Eq.(\ref{effm})).
The mode functions satisfy the equivalent integral equation
\begin{equation} 
U_k(t)=e^{-i\omega_{k0} t}+
\int\limits^{\infty}_{0}\!{\rm d}t'
\Delta_{k,{\rm ret}}(t-t'){\cal V}(t')U_k(t')\;.
\end{equation}
with
\begin{equation} 
\label{fvt}
\Delta_{k,{\rm ret}}(t-t')= -\frac{1}{\omega_{k0}}
\theta(t-t')\sin\left(\omega_{k0}(t-t')\right) \; .
\end{equation}
For $U_k(t)$ we choose the following ansatz
\begin{equation} 
\label{ansatz}
U_k(t)=e^{-i\omega_{k0} t}(1+f_k(t)) \; ,
\end{equation}
to separate $U_k(t)$ into the trivial part corresponding to the case
${\cal V}(t)=0$ and a function $f_k(t)$ which represents the reaction
to the potential.
$f_k(t)$ satisfies the differential equation
\begin{equation} \label{fdiffeq}
\ddot{f}_k(t)-2i\omega_{k0}\dot{f}_k(t)=-{\cal V}(t)(1+f_k(t))\;,
\end{equation}
with the initial conditions $f_k(0)=\dot{f}_k(0)=0$ or the equivalent
integral equation
\begin{equation} \label{finteq}
f_k(t)=\int\limits^{t}_{0}\!{\rm d}t'\Delta_{k,{\rm ret}}
(t-t'){\cal V}(t')[1+f_k(t')]e^{i\omega_{k0} (t-t')}\;.
\end{equation}
We expand now $f_k(t)$ with respect to orders in ${\cal V}(t)$
by writing
\begin{eqnarray}
\label{entwicklung}
f_k(t)&=& f_k^{(1)}(t)+f_k^{(2)}(t)+f_k^{(3)}(t) +\cdots \\
 &=& f_k^{(1)}(t)+f_k^{\overline{(2)}}(t)
\; ,\end{eqnarray}
where $f_k^{(n)}(t)$ is of $n$th order in ${\cal V}(t)$ and 
$f_k^{\overline{(n)}}(t)$
is the sum over all orders beginning with the $n$th one:
\begin{equation} 
f_k^{\overline{(n)}}(t)=\sum_{l=n}^\infty f_k^{(n)}(t)
\; .\end{equation}
The function $f_k^{\overline{(1)}}(t)$ is
identical to the function $f_k(t)$ itself which is obtained
by solving (\ref{fdiffeq}). The function
$f_k^{\overline{(2)}}(t)$ can be computed
by using the differential equation,
via
\begin{equation} \label{f2diffeq}
\ddot{f}_k^{\overline{(2)}} (t)-2i\omega_{k0}
\dot{f}_k^{\overline{(2)}}(t)=-{\cal V}(t)f_k^{\overline{(1)}}(t) \; ,
\end{equation}
or by iteration via
\begin{equation} \label{f2inteq}
f_k^{\overline{(2)}}(t)=
\int\limits^{t}_{0}\!{\rm d}t'\Delta_{k,{\rm ret}}
(t-t'){\cal V}(t')
f_k^{\overline{(1)}}(t')e^{i\omega_{k0} (t-t')} \;.
\end{equation}

This iteration has the advantage for the numerical computation 
that it avoids computing
$f_k^{\overline{(2)}}$ via the small difference
$f_k^{\overline{(1)}}-f_k^{(1)}$. However, the integral equations
are used as well in order to derive the asymptotic behaviour as
$\omega_{k0}\to \infty$ and to separate divergent and finite 
contributions. The leading orders of $f_k(t)$ are discussed in detail
in \cite{Baacke:1997c,Baacke:1997a,Baacke:1997b} at full length and
we do not want to repeat it here. In this work we are more interested
in the effects of the finite contributions at finite temperature
and in the self consistent solving of the large-$N$ limit.


\section{Renormalization}
\setcounter{equation}{0}

We use the expansion and the definition introduced in the
previous section in order to
in order to single out the divergent
terms from the fluctuation integral; we have
\bea \label{lNmassexp}
\calm^2(t)&=&m^2+\delm^2+ 
(\lambda+\dellam)\left\{\phi^2(t)+\imone(m_0,T)
\right. \\ \nonumber && \left.-\imthree(m_0,T)
\left[\calm^2(t)-\calm^2(0)\right]+\calff(t,T)\right\}\kma
\eea
where the finite part of $\calf(t,T)$ can be written as
\bea
\calff(t,T)
 &=&\intk
\frac{1}{2\omega_{k0}^2}\int\limits^{t}_{0}\!
{d}t'\cos\left(2\om(t-t')\right )\dot{\calv}(t')
\coth{\frac{\beta\om}{2}} \nonumber\\
&+&\intk\left\{2 {\rm 
Re}f_k^{\overline {(2)}}(t)
+|f_k^{\overline {(1)}}(t)|^2\right\}\coth{\frac{\beta\om}{2}} \kma
\eea
and where the divergent integrals are defined as
\bea
\label{imone}
\imone(m_0,T)&=&\intk\left(1+\frac{2}{e^{\beta\omega_0}-1}\right)
=I_{-1}(m_0)+\Sigma_{-1}(m_0,T)\\
\label{imthree}
\imthree(m_0,T)&=&\int\frac{d^3k}{(2\pi)^3 4\om^3}
\left(1
+\frac{2}{e^{\beta\omega_0}-1}\right) 
=I_{-3}(m_0)+\Sigma_{-3}(m_0,T)
\pkt
\eea
The integrals $I_{-k}(m_0)$ are those which occur in the
renormalization at $T=0$.
Their dimensionally regularized form will be given
below. The additional temperature dependent terms $\Sigma_{-k}(m_0,T)$
are finite. They are defined as
\bea
\Sigma_{-1}(m_0,T)&=&
\intko\frac{1}{\om\left(e^{\beta\om}-1\right)}\\
\Sigma_{-3}(m_0,T)&=&
\intko\frac{1}{2\om^3\left(e^{\beta\om}-1\right)} \pkt
\eea
We derive some useful explicit expressions for these
integrals in Appendix A.

It is convenient to include these
finite terms into the definition of $\calff(t,T)$.
Then the time dependent mass takes the form
\begin{equation}
\label{mass}
{\cal M}^2(t)=m^2+\delta m^2+(\lambda+\dellam)\left[
\phi^2(t)+\imone(m_0)-\imthree(m_0){\cal V}(t)+\calfft(t,T)\right]
\kma
\end{equation}
with
\begin{equation}
\calfft(t,T)=\Sigma_{-1}(m_0,T)-{\cal V}(t)
\Sigma_{-3}(m_0,T)+
{\cal F}_{\rm fin}(t,T)\pkt
\end{equation}
The time dependent mass (\ref{mass}) contains both renormalization
constants $\delm$ and $\dellam$. Furthermore, its definition by this
equation is implicit, $\calm^2(t)$ appears also on the right
hand side of Eq. (\ref{mass}) in $\calv(t)$. 

We now set out to fix the renormalization counter terms in such
a way that the relation between  the time-dependent mass 
and $\phi(t)$ becomes finite.
An additional constraint derives from the requirement
that the  renormalization
counter terms should not depend on the initial condition but only
on the parameters appearing in the Lagrangian, i.e.,
$\lambda$ and $m$. For the simpler case
of the one-loop equations  this has been 
achieved \cite{Baacke:1997a}.

  We first determine $\dellam$ by considering
the difference
\bea
\calv(t)&=&\calm^2(t)-\calm^2(0) \\ \nonumber 
&=&(\lambda+\dellam)\left\{\phi^2(t)-\phi^2(0)-\imthree(m_0)
\calv(t)+\calfft(t,T)-\calfft(0,T)\right\}
\eea
or
\bea  \label{calvimp}
\calv(t)\left[1+(\lambda+\dellam)\imthree(m)\right]&=&
(\lambda+\dellam)\left\{\phi^2(t)-\phi^2(0) 
-\left[\imthree(m_0)-\imthree(m)\right]
\calv(t)\right.\nonumber\\&&\left.+\calfft(t,T)-
\calfft(0,T)\right\}\pkt \eea
We now require
\be 
\label{dellamcond}
\frac{\lambda+\dellam}{1+(\lambda+\dellam)\imthree(m)}=\lambda\pkt
\ee
Solving with respect to $\dellam$ we find
\be \label{dellam}
\dellam= \frac{\lambda^2\imthree(m)}
{1-\lambda\imthree(m)}\pkt
\ee
Inserting this relation into (\ref{calvimp}) we have
\be  \label{calvimp2}
\calv(t)=
\lambda\left\{\phi^2(t)-\phi^2(0)-
\left[\imthree(m_0)-\imthree(m)\right]\calv(t)
+\calfft(t,T)-\calfft(0,T)\right\}
\ee
or
\be  \label{calvexp}
\calv(t)=
\frac{\lambda}{1+\lambda\left[\imthree(m_0)-\imthree(m)\right]}
\left[\phi^2(t)-\phi^2(0)
+\calfft(t,T)-\calfft(0,T)\right]\pkt
\ee
This is a finite relation for the potential $\calv(t)$
since the difference $[\imthree(m_0)-\imthree(m)]$ is finite.
Using dimensional regularization 
\be
\imthree(m_0)=\left\{\intko \frac{1}{4\om^3}
\right\}_{\rm{reg}}=
 \frac{ 1}
{16\pi^2}\left\{\frac{2}{\epsilon}
+\ln{\frac{4\pi\mu^2}{m_0^2}}-\gamma\right\}
\kma \ee
and therefore
\be
\imthree(m_0)-\imthree(m)=
\frac{1}{16\pi^2}\ln\left(\frac{m^2}{m_0^2}\right)\pkt
\ee
We now go back to equation (\ref{lNmassexp}) which we take at
the initial time $t=0$:
\be\label{lNmass0}
m_0^2\equiv\calm^2(0)=m^2+\delm^2+(\lambda+\dellam)
\left[\phi^2(0)+\imone(m_0)+\calfft(0,T)\right]\pkt
\ee
This is an implicit relation between $m_0$ and $\phi(0)$ which, however,
contains still the infinite quantities $\dellam,\delm$ and
$\imone(m_0)$. In order to proceed we note the following
explicit relation between $\imone$ and $\imthree$ which follows
from the dimensionally regularized expressions for these
quantities
\bea
\left\{\intk\right\}_{\rm{reg}}
&=&-\frac{ m_0^2}
{16\pi^2}\left\{\frac{2}{\epsilon}+
\ln{\frac{4\pi\mu^2}{m_0^2}}-\gamma+1\right\}\nonumber\\
&=&
-m_0^2 \imthree(m_0)-\frac{m_0^2}{16\pi^2}\pkt
\eea
Therefore we can rewrite (\ref{lNmass0}) as
\be
m_0^2=m^2+\delm^2+(\lambda+\dellam)
\left[\phi^2(0)-m_0^2\imthree(m_0)-
\frac{m_0^2}{16\pi^2}+\calfft(0,T)\right]
\ee
or
\bea
&&m_0^2\left[1+(\lambda+\dellam)\imthree(m)\right]= 
m^2+\delm^2 \\ &&\hspace{10mm} \nonumber+(\lambda+\dellam)
\left[\phi^2(0)-m_0^2(\imthree(m_0)-\imthree(m))-
\frac{m_0^2}{16\pi^2}+\calfft(0,T)\right]\pkt
\eea
We now require the factors $\lambda+\dellam$ and
$[1+(\lambda+\dellam)\imthree]$ to cancel on account of
(\ref{dellamcond}), so as to
obtain a relation between finite quantities.
This is obviously the case if
 \be \label{delmcond} 
m^2+\delm^2=m^2\left[1+(\lambda+\dellam)\imthree(m)\right]
+(\lambda+\dellam)\rho
\ee
since then
\be \label{gapeq}
m_0^2-m^2=\lambda
\left\{\phi^2(0)-m_0^2\left[\imthree(m_0)-\imthree(m)\right]-
\frac{m_0^2}{16\pi^2}+\rho +\calfft(0,T)\right\}\pkt
\ee
(\ref{delmcond}) fixes $\delm^2$ as
\be
\delm^2=(\lambda+\dellam)
\left[m^2\imthree(m)+\rho\right]=
\lambda\frac{m^2\imthree(m)+\rho}{1-\lambda 
\imthree(m)}\pkt
\ee
Setting $\rho=0$ we have
\be
\delm^2=\frac{\lambda m^2\imthree(m)}{1-\lambda 
\imthree(m)}\kma
\ee
and
\be \label{gapeq1}
m_0^2-m^2=\lambda
\left\{\phi^2(0)-m_0^2\left[\imthree(m_0)-\imthree(m)\right]-
\frac{m_0^2}{16\pi^2}+\calfft(0,T)\right\}\pkt
\ee
This corresponds to the $\overline{MS}$ subtraction.
Another natural choice is $\rho=m^2/16\pi^2$; then
\be
\delm^2=\lambda
\frac{ m^2\imthree(m)+m^2/16\pi^2}{1-\lambda 
\imthree(m)}=-\frac{\lambda \imone(m)}{1-\lambda 
\imthree(m)}\kma
\ee
and
\be \label{gapeq2}
m_0^2-m^2=\lambda
\left\{\phi^2(0)-m_0^2\left[\imthree(m_0)-\imthree(m)\right]-
\frac{m_0^2-m^2}{16\pi^2}+\calfft(0,T)\right\}\pkt
\ee
This choice is analogous to the one 
in \cite{Baacke:1997a} for the one-loop
equations and will be used in the following.

The ``gap equation'' (\ref{gapeq}) and the renormalized 
definition of the potential (\ref{calvexp}) constitute, along
with the equations of motion the
basic renormalized equations for the self consistent large-$N$ dynamics.

The gap equation has to be solved at $t=0$ and determines the
relation between $m_0$ and $\phi(0)$. For later times we have
\bea
\calm^2(t)&=&m_0^2+\calv(t) \\ \nonumber
&=&m_0^2+\cc\lambda\left[\phi^2(t)-\phi^2(0)
+\calfft(t,T)-\calfft(0,T)\right]\kma
\eea
with
\be
\cc=\left(1+\frac{\lambda}{16\pi^2}\ln{\frac{m^2}{m_0^2}}\right)^{-1}
\pkt\ee
Since the gap equation can be cast into different forms we
can obtain several equivalent forms of this equation. 
Solving the gap equation for $\phi^2(0)$ we find
\be
-\lambda\cc
\phi^2(0)=-m_0^2
+\cc
\left[m^2-\lambda
\left(\frac{m_0^2}{16\pi^2}-\rho-\calfft(0,T)\right)\right]\kma
\ee
so that 
\be \label{massfin}
\calm^2(t)=
\cc
\left[m^2+\lambda
\left(\phi^2(t)-\frac{m_0^2}{16\pi^2}+\rho
+\calfft(t,T)\right)\right]\pkt
\ee
Having obtained a finite relation between $\phi(t)$ and $\calm(t)$ the 
equations of motion for the classical field $\phi(t)$ and for the
modes $U_k(t)$ are well-defined and finite.

Here we have chosen to include the
corrections of leading order, proportional to 
$\Sigma_{-1}(m_0,T)$, into the finite part of the fluctuation integral.
These terms are important at high temperature; they appear in the
gap equation via $\calfft(0,T)=\Sigma_{-1}(m_0,T)\simeq 
T^2/12$. Omitting some terms of order  $\lambda/16\pi^2$
the gap equation (\ref{gapeq}) becomes
\be
m_0^2 \simeq m^2+\lambda\phi(0)^2 +\frac{\lambda}{12} T^2 \pkt
\ee
Therefore, at high temperature the mass circulating 
in the loop is dominated by the ``hard'' $\lambda T^2$ term.

We will need in the following the fluctuation integral $\calf(t,T)$
which is and will remain divergent. We need an expression in which 
these divergencies appear in explicit form. We use
\be
\calf(t,T)=I_{-1}(m_0)-I_{-3}(m_0)\left[\calm^2(t)-\calm^2(0)\right]
+\calfft(t,T)
\ee
and insert the expression for $\calm^2(t)$ we have just derived.
Using the gap equation and some reshuffling of terms
we obtain
\bea \label{calfex}
\calf(t,T)&=&-\frac{m_0^2}{16\pi^2}-\cc\lambda I_{-3}(m_0)
\phi^2(t)-m^2\cc I_{-3}(m_0)\\ \nonumber
&&+\cc\frac{\lambda}{16\pi^2}
I_{-3}(m_0)(m_0^2-m^2)+\cc(1-\lambda I_{-3}(m))\calfft(t,T)
\pkt \eea


\section{Renormalization of energy and pressure}
\setcounter{equation}{0}

The expressions for the energy density and for the pressure
have been given in section 2. Apart from the renormalization
counter terms which we have already fixed in renormalizing
the equation of motion, two new counter terms appear,
the ``cosmological constant''   term $\delta\Lambda$ in the
energy density and the ``improvement term''
$A d^2( \phi^2+\langle \psi^2\rangle)/dt^2$ in the pressure.
These terms must suffice for rendering the expressions for
energy density and pressure finite.

We start with the expression (\ref{energy}) for the energy
which we rewrite as
\bea
\calE &=&\frac{1}{2}\dot \phi^2(t)+\frac{1}{2}m^2\phi^2(t) 
+\frac{\lambda}{4}\phi^4(t)\nonumber\\&& +\calE_{\rm fl}(t,T)
-\frac{\lambda+\dellam}{4}\calf^2(t,T) 
+\frac{1}{2}\delm^2\phi^2(t)+\frac{\dellam}{4}\phi^4(t)
+\delta\Lambda \pkt
\eea
with
\be
\calE_{\rm fl}(t,T)=\intk
\coth{\frac{\beta\om}{2}} \left\{\frac 1 2|\dot{U}_k(t)|^2
+\frac{1}{2}\omega^2_k(t)|U_k(t)|^2\right\}
\pkt
\ee 
In the latter expression we split the Bose factor as before
\be \label{split}
\coth{\frac{\beta\om}{2}}=1+\frac{2}{e^{\beta\om}-1}
\pkt\ee
The integrations involving the second term are finite, we define
\be \label{eflT} 
\Delta\calE_{\rm fl}(t,T)=\intk
\frac{2}{e^{\beta\om}-1} \left\{ \frac 1 2|\dot{U}_k(t)|^2
+\frac{1}{2}\omega^2_k(t)|U_k(t)|^2\right\}
\pkt\ee
Those involving the first term have been discussed 
in \cite{Baacke:1997a}. Following this discussion we can 
decompose the integral via
\be 
\calE_{\rm fl}(t,0)=I_1(m_0)+\frac{1}{2}\calv(t)I_{-1}(m_0)
-\frac{1}{4}\calv^2(t)I_{-3}(m_0)+
\calE_{\rm fl,fin}(t,0)
\ee
with \footnote{We have overlooked in \cite{Baacke:1997a}
that in the expression for the fluctuation energy
given there two terms cancel on account of a Wronskian identity,
given in \cite{Baacke:1997c}, Eq. (73). They did so, of course, in the
numerical calculations.}
\be \label{efl0}
\calE_{\rm fl,fin}(t,0)
=
\frac{1}{2}\intk \left\{\frac{1}{2}|\dot f^{\overline{(1)}}_k|^2
+\frac{\calv(t)}{2}\left[2\re f_k^{\overline{(1)}}
+|f^{\overline{(1)}}_k|^2\right]+\frac{\calv^2(t)}{8\om^2}\right\}
\pkt\ee
We denote the sum of both
finite contributions as $\calE_{\rm fl,fin}(t,T)$. The expression for
the energy now takes the form
\bea \label{ediv2}
\calE &=&\frac{1}{2}\dot \phi^2+\frac{1}{2}m^2\phi^2 
+\frac{\lambda}{4}\phi^4 +\calE_{\rm fl,fin}(t,T)
+I_1(m_0)+\frac{1}{2}\calv(t)I_{-1}(m_0)
-\frac{1}{4}\calv^2(t)I_{-3}(m_0)  \nonumber \\
&&-\frac{\lambda+\dellam}{4}\calf^2(t,T) 
+\frac{1}{2}\delm^2\phi^2+\frac{\dellam}{4}\phi^4
+\delta\Lambda \pkt
\eea
In addition to the divergent integrals $I_n(m_0)$ and the
counter terms further divergencies are contained in the 
fluctuation integral $\calf(t,T)$; these are given explicitly in
Eq. (\ref{calfex}).
The analysis of  Eq. (\ref{ediv2}), after inserting the explicit
expressions for $\calv(t)$ and $\calf(t,T)$, becomes rather cumbersome.
To give an outline of the typical algebraic manipulations we consider
explicitly the coefficents of $\phi^4$. Collecting everything except
the bare $\lambda\phi^4$ term we find that $\phi^4(t)$ is
multiplied by a sum of divergent terms
\be
\frac{\dellam}{4}-\frac{\lambda+\dellam}{4}\cc^2\lambda^2x_0^2-
\frac{1}{4}\cc^2x_0\lambda^2 \pkt
\ee
We use the abbreviations $x_0=I_{-3}(m_0)$ and $x=I_{-3}(m)$, so that
\be
\cc=\frac{1}{1-\lambda (x_0-x)}
\ee
and (see Eq. (\ref{dellam}))
\be
\dellam=\frac{\lambda^2}{1-\lambda x}\pkt
\ee
One finds that all divergent quantities 
combine into the finite expression 
\be
-\frac{1}{4}\lambda^2\cc(x-x_0)=\frac{\lambda^2}{64\pi^2}\cc
\ln \left(\frac{m^2}{m_0^2}\right)\equiv\frac{\Delta\lambda}{4} \kma
\ee
so the correction to the $\phi^4$ term in the energy
becomes
\be
\frac{\Delta\lambda}{4}\phi^4\pkt
\ee
Collecting similarly all terms proportional to
$\phi^2$ one finds that the correction to the mass term becomes
finite as well, explicitly
\be
\frac{1}{2}\Delta m^2\phi^2=\frac{\lambda}{32\pi^2}\cc \left[
m^2-m_0^2-m^2\ln\left(\frac{m^2}{m^2_0}\right)\right]\phi^2\pkt
\ee
 There are further time-dependent terms proportional to
$\calfft^2(t,T)$ and $\calfft(t,T)$ and constant terms. The 
divergent parts of the latter ones can be absorbed into 
$\delta\Lambda$, the coefficient of the term linear in
$\calfft$ vanishes and the quadratic one has a finite coefficient.
The counter term $\delta \Lambda$ can be chosen independent
of $m_0$:
\be
\delta \Lambda=\frac{m^4}{4(1-\lambda x)}
\left(x+\frac{1}{8\pi^2}
-\frac{\lambda}{256 \pi^4}\right),
\ee
there remains a finite constant
\be
\Delta\Lambda=
\frac{1}{4}\cc\left[(x_0-x)m^4+\frac{1}{8\pi^2}(m^2-m_0^2)+
\frac{1}{32\pi^2}m_0^4+\frac{\lambda}{256\pi^4}(m_0^2-m^2)^2\right]
\pkt
\ee
So the expression for the energy can really be rendered finite with 
counter terms independent of the initial condition.
Explicitly we find
\bea \label{ediv}
\calE &=&\frac{1}{2}\dot \phi^2+
\frac{1}{2}(m^2+\Delta m^2)\phi^2 
+\frac{\lambda+\Delta\lambda}{4}\phi^4 
\\ \nonumber&&+\calE_{\rm fl,f}(t,T)
-\frac{\lambda}{4}\cc\calfft^2(t,T)+\Delta\Lambda
\pkt\eea
We finally have to give a finite expression for the
pressure, using our last free counter term.
We write the pressure in the form
\be
p=\dot\phi^2(t)-\calE+p_{\rm fl}(t,T)+
A\frac{d^2}{dt^2}\left[\phi^2(t)+\calf(t,T)\right]\pkt
\ee
Here we have anticipated a special form of the counter term, indeed
for the expression in brackets one can choose a priori
an arbitrary Lorentz
scalar, the additional piece of the energy momentum tensor being
trivially conserved on account of its tensor structure
$\partial_\mu\partial_\nu-g_{\mu\nu}\partial^2$. Of course  it has
to be suited for the renormalization procedure.
The fluctuation part of the pressure consists again of three parts,
a divergent one, a finite one independent of the temperature and
a finite integral involving the thermal distribution function
$1/(\exp(\om/T)-1)$. The analysis for $T=0$ has been performed
in \cite{Baacke:1997a}. Following the discussion there
we can write $p_{\rm fl}$ as
\be
p_{\rm fl}(t,T)=p_{\rm fl,fin}(t,0)+\Delta p_{\rm,fl}(t,T)
-\frac{m_0^4}{96\pi^2}-\frac{m_0^2}{48\pi^2}\calv(t)
-\frac{1}{6}(I_{-3}(m_0)+\frac{1}{48\pi^2})\ddot\calv(t)\pkt
\ee
$\Delta p_{\rm,fl}(t,T)$ is given by
\be
\Delta p_{\rm,fl}(t,T)=\intk\frac{2}{e^{\beta\om}-1}
\left(\om^2+\frac{k^2}{3}\right)|U_k(t)|^2 \kma
\ee
the $T=0$ finite part by
\bea
p_{\rm fl,fin}(t,0)&=&
\intk
\left\{\left(\om^2+\frac{\vec k^2}{3}\right)
\left[2\re f_k^{\overline{(2)}}(t)+|f_k^{\overline{(1)}}(t)|^2
\right]\right.\nonumber\\
&&
+\left(\frac{1}{6\om^2}-\frac{m_0^2}{24\om^4}\right)
\int\limits_0^{t}\!dt\,
\cos{2\omega_k^0(t-t')}\tdot{\calv}(t')\nonumber\\
&&+\left(\frac{1}{12\om^2}+\frac{m_0^2}{24\om^4}\right)
\cos(2\om t)\ddot{\calv}(0)\\
&&+|\dot{f}_k^{\overline{(1)}}(t)|^2
-2\re\left[i\om\dot{f}^{\overline{(1)}}(t)
+i\om f_k^{\overline{(1)}}(t)
f_k^{\overline{(1)}*}(t)\right]
\biggr\}\; .
\eea
We call the sum of both
finite fluctuation integrals $p_{\rm fl,fin}(t,T)$. 
Now we have to consider the divergent terms.
We observe that $\ddot\calv(t)$ is given by
\be \label{vddot}
\ddot\calv(t)=\lambda\cc 
\frac{d^2}{dt^2}\left[\phi^2(t)+\calfft(t,T)\right]
\pkt\ee
On the other hand, using Eq. (\ref{calfex}) we have 
\be
\frac{d^2}{dt^2}\calf(t,T)=
\frac{d^2}{dt^2}\left[-\lambda \cc x_0 \phi^2(t)+
\cc(1-\lambda x)\calfft(t,T)\right]
\ee
and therefore
\be
A\frac{d^2}{dt^2}(\phi^2(t)+\calf(t,T))=
A\frac{d^2}{dt^2}\cc (1-\lambda x)\left[\phi^2(t)+\calfft(t,T)\right]
\ee
As apparent from Eq. (\ref{vddot}) this 
matches in form with the divergent term
$I_{-3}(m_0)\ddot\calv(t)/6$.
Insisting again in choosing the counter term independent of the initial
condition we fix
\be
A=\frac{\lambda x}{6(1-\lambda x)}=\frac{\lambda I_{-3}(m)}
{6(1-\lambda I_{-3}(m))}
\ee
and retain a finite term 
\be
-\frac{1}{96\pi^2}\left[\ln\left(\frac{m^2}{m_0^2}\right)+2
\right]\ddot \calv(t)\pkt
\ee
The final result for the pressure reads
\be
p=\dot\phi^2(t)-\calE+p_{\rm fl,fin}(t,T)
-\frac{m_0^4}{96\pi^2}-\frac{m_0^2}{48\pi^2}\calv(t)
-\frac{1}{96\pi^2}\left[\ln\left(\frac{m^2}{m_0^2}\right)+2
\right]\ddot \calv(t)\pkt
\ee

A further quantity of interest is the particle number density. 
It does not need to be renormalized after subtraction
of the initial particle number density and is given by
\bea
n(t)-n(0) &=& 
\intko
\coth{\frac{\beta\om}{2}}
 \left\{\frac{1}{4}\left[|U_k(t)|^2+\frac{1}{\om^2}
|\dot U_k(t)|^2\right]-\frac{1}{2}\right\}
\nonumber\\
&=&\intko
\coth{\frac{\beta\om}{2}}
\frac{ |\dot f^{\overline{(1)}}(t)|^2}{4\om^2} \pkt
\eea


\section{Numerical results}

We have implemented numerically the renormalized 
formalism derived in the previous sections. The
results show essentially the same features as those
found by other groups \cite{Boyanovsky:1997e,Cooper:1997}.

We have chosen several parameter sets which are
displayed in Table 1.
All parameter sets start with an initial value of the
mass $m_0=3$. The gap equation is then solved for 
$\phi(0)$. We have chosen two parameter sets at $T=0$,
one with $\lambda=1$ and one with $\lambda=5$. Another
two parameter sets have finite temperature, $T=3$ and
$T=10$, respectively, and $\lambda=1$. Parameter sets
with smaller temperatures showed very little deviations
from the $T=0$ case and are not presented.

For the first parameter set the numerical results 
are diplayed in Figs. 1 a-e. The classical field
$\phi(t)$ is seen to oscillate with an amplitude
decreasing slowly to an asymptotic value. The 
potential $\calv(t)$, the difference between $\calm^2(t)$
and $m_0^2=\calm^2(0)$ reaches an asymptotic average
of $-4.0$ with small oscillations. Classical and fluctuation 
energy are shown in Fig. 1c. We denote as fluctuation energy
the quantity $\calE_{\rm fl,fin}(t,T)=\calE_{\rm fl,fin}(t,0)
+\Delta \calE_{\rm fl,fin}(t,T)$, see Eqs. (\ref{efl0},\ref{eflT}).
The remaining parts of the total energy (\ref{ediv}) are 
considered as the classical energy. The separation is somewhat
arbitrary as finite parts of the leading order fluctuation energy
are contained in $\Delta\lambda,\Delta m^2$ and $\Delta \Lambda$.
So the fact that the ``classical energy'' becomes even
negative is deceptive, the classical amplitude $\phi(t)$
has not decreased to zero. It has decreased roughly by a factor
of 3, implying a decrease of energy  by a factor 9.
Nevertheless the production of fluctuation energy
is important, as also seen from the particle number
displayed in Fig. 1e.
The energy is seen to be conserved, apart from some numerical
noise in the early stage of evolution, due to badly 
convergent integrals. The pressure is seen to approach an
asymptotic value of $\simeq 5$, somewhat smaller than
the ultrarelativistic limit of $E/3=6.7$. 

For the other parameter sets we display just the classical
amplitude and the potential $\calv(t)$, need to verify a 
sum rule (see below). For the finite temperature case
$T=3$ the particle number develops almost identically
to the one for parameter set 1, and reaches an asymptotic value
of $\simeq 8.5$. This can be compared with the
thermal particle number density $3.29$. 
The situation changes strongly in the high temperature situation,
parameter set 4.
There, the particle production, displayed in Fig. 4c, is insignificant
with respect to the thermal particle number density which is
$n(0)\simeq 122$. The other Figures for this parameter set show clearly
that the interaction of the classical amplitude with
fluctuations is suppressed. 

An interesting topic which has emerged recently
\cite{Boyanovsky:1997e} is a sum rule for the
late time behaviour, which was found to be satisfied
numerically with high precision.
Adapted to our notation and definitions, and generalized
(naively) to finite temperature, it reads
\be \label{sumrule}
\calm^2(\infty)\simeq\cc\left(m^2-\frac{\lambda}{16\pi^2}(m_0^2-m^2)
+\Sigma_{-1}(m_0,T)\right)+\lambda \cc \frac{\phi^2(0)}{2}
\pkt
\ee 
This value of $\calm(\infty)$ is determined by the lower limit
of a parametric resonance band, essentially it implies that
the classical oscillation becomes stationary if its frequency,
i.e. $\calm(t)$, settles in such a way as to
avoid resonant excitation. 
We have verified this sum rule for our parameter sets, the
left and right hand sides of the sum rule are compared in Table 1.
For $T=0$ the agreement is excellent, for $T\neq0$, a case
not considered in \cite{Boyanovsky:1997e}, the deviations
are of the order of $10 \%$.


\section{Conclusions}

We have derived in this paper the renormalized equations
of motion, the energy and the pressure for the
nonequilibrium evolution of a scalar $O(N)$ model.
The regularization has been done in a covariant way, using
dimensional regularization. As in the case of the
one-loop equations studied previously \cite{Baacke:1997a} it
was possible to fix all the renormalization counter terms 
independent of the initial conditions, though the divergent
integrals appearing in the unrenormalized expressions do
depend on the ``initial mass'' $m_0$ instead of the renormalized mass.
As a renormalization convention we have chosen a slightly
modified $\overline{MS}$ scheme, it can of course be 
modified to another suitable convention like 
renormalization at the minimum of the effective potential.

We have restricted the formalism to the case of
unbroken symmetry. The special - and highly interesting -
aspects of the broken symmetry case have been investigated
in \cite{Cooper:1997}. Here it was our aim to present a 
general framework, which has to be adapted to specific
physical models. In the case of the one-loop equations
it was indeed possible to extend the formalism to 
nonabelian gauge theories \cite{Baacke:1997b}.

The formalism developed here represents at the same time
a rather convenient computation scheme. The 
CPU time requirements
are of the same order as the one for the one-loop
equations. Typically, the examples we have presented
took 1-2 hours each on a small workstation. We 
have not attempted to meet the same standards in numerical
precision as other groups, nevertheless our results
show the same general features as those of other groups.
This is presumably due to the fact that the system
as such has a stable and essentially predictable
late time behavior \cite{Boyanovsky:1997e}, indeed our results
fulfil an asymptotic sum rule formulated in this Reference.

\begin{appendix}


\section{Some thermal integrals}
\setcounter{equation}{0}
In this Appendix we give, without claim of originality,
 some explicit
expressions for the thermal integrals as we have used them in the
numerical computations. In deriving these relations we have relied
on the integral tables of Prudnikov, Brychkov and
Marichev \cite{Prudnikov:1986}.

The finite temperature part of the 
tadpole graph, which constitutes
a correction to the mass, is 
given by the integral 
\bea \label{sm1ex}
&&\Sigma_{-1}(m_0,T)=
\intko\frac{1}{\omega_{k0}(e^{\beta\omega_{k0}}-1)}\\
&&=\frac{m_0^2}{2\pi^2}\sum_{n=1}^{\infty}
\left\{\tnm^{-2}
+\sum_{j=0}^\infty\frac{1}{4j!(j+1)!}\left[2\ln{\frac{\tnm}{2}}
-\psi(j+1)-\psi(j+2)\right]\left(\frac{\tnm^2}{4}\right)^j
\right\}\nonumber
\eea
where $\tnm$ stands for $n m_0/T$.
For large values of $\tnm$ (this means for small $T$)
the  integrand is dominated by momenta of order
$k \simeq T$. Therefore one can expand $\om$ w.r.t.
powers of $m/k$; the integral is then well approximated by
\be
\label{sm1app}
\Sigma_{-1}(m_0,T)\simeq
\frac{m_0^2}{2\pi^2}\sum_{n=1}^{\infty}\sqrt{\frac{\pi}{2}}e^{-\tnm}
\tnm^{-\frac 3 2}\left\{1+\frac{3}{8}
\tnm^{-1}-\frac{15}{128}\tnm^{-2}
+\frac{105}{1024}\tnm^{-3}
+{\cal O}\left(\tnm^{-4}\right)\right\}
\pkt
\ee
For $T \gg m_0$ we find directly from Eq. (\ref{sm1ex})
the well-known
approximation
\be \label{sm1ht}
\Sigma_{-1}(m_0,T)\simeq
\frac{1}{2\pi^2}\zeta(2) T^2=\frac{T^2}{12}\pkt
\ee
It yields the hard thermal loop corrections to the mass.

The finite temperature part of the fish graph,
which can be considered as a finite 
correction to the coupling constant,
is given by
\bea \label{sm3ex}
\Sigma_{-3}(m_0,T)&=&\intko\frac{1}{2\omega_{k0}^3}\frac{1}
{e^{\beta\omega_{k0}}-1}\\
&=&\frac{1}{4\pi^2}\sum_{n=1}^\infty\sum_{j=0}^\infty
\left\{\frac{1}{(2j-1)j!j!}
\left[\ln{\frac{\tnm}{2}}-\psi(j+1)-\frac{1}{2j-1}\right]
\left(\frac{\tnm^2}{4}\right)^j+\frac{\pi}{2}\tnm\right\}
\nonumber
\eea
For small $T$ or large  $\tnm$ we find the approximation
\bea
\label{sm3app}
\Sigma_{-3}(m_0,T)
\simeq-\frac{1}{4\pi^2}\sum_{n=1}^{\infty}{\frac{\sqrt\pi}{2}}e^{-\tnm}
\tnm^{-\frac 3 2}
\left\{1-\frac{21}{8}
\tnm^{-1}-\frac{1185}{128}\tnm^{-2}
-\frac{42735}{1024}\tnm^{-3}
+{\cal O}\left(\tnm^{-4}\right)\right\}
\pkt
\eea
For large temperatures this integral behaves linear in $T$, more precisely
\be
\label{sm3ht}
\Sigma_{-3}(m_0,T)\simeq \frac{1}{8\pi}\frac{T}{m} 
\pkt \ee
The finite temperature part associated with the 
quartic divergence in the energy
is given by the Planck formula
\begin{eqnarray}
\Sigma_1(m_0,T)&=&
\intko\frac{\omega_{k0}}{e^{\beta\omega_{k0}}-1}\nonumber\\
&=&
\frac{m_0^4}{2\pi^2 }\sum_{n=1}^{\infty}
\left\{6\tnm^{-4}\left(1-\frac{\tnm}{12}
\right)\right.\\
&&+\frac{1}{16}\sum_{j=0}^{\infty}\frac{2j+1}{j!(3)_j}\left(
2\ln{\frac{\tnm}{2}}-\psi(3+j)-\psi(1+j)+\frac{2}{2j+1}
\right)\biggr\}\nonumber\pkt
\end{eqnarray}
As an approximation for large $\tnm$ or small $T$ we find
\be
\label{approxen}
\Sigma_1\simeq\frac{m_0^4}{2\pi }\sum_{n=1}^\infty e^{-\tnm}
\sqrt{\frac{\pi}{2}}\tnm^{-\frac 3 2}
\left\{1+\frac{27}{8}\tnm^{-1}
+\frac{705}{128}\tnm^{-2}
+\frac{2625}{1024}\tnm^{-3}
+{\cal O}\left(\tnm^{-4}\right)\right\}\pkt
\ee
For large temperatures one obtains
\be
\Sigma_1(m_0,T) \simeq \frac{\pi^2}{30} T^4
\pkt\ee

\end{appendix}


\newpage
\section*{Table Captions}
{\bf Table 1:} Parameter sets and sum rule. We display the 
parameters of the 4 numerical simulations. The mass unit is 
set by $m=1$; $\phi(0)$ follows from the gap equation
(\ref{gapeq}), $\calm^2(\infty)$ is the sum of $m_0^2$
and of $\calv(\infty)$ read from the corresponding Figures.
R.h.s. is the right hand side of Eq.(\ref{sumrule}).
\\

\section*{Figure Captions}
\noindent
{\bf Fig. 1a:} $\phi(t)$ for parameter set 1. \\
{\bf Fig. 1b:} The potential $\calv(t)$ for parameter set 1. \\
{\bf Fig. 1c:} Classical energy (long-dashed line), fluctuation energy 
$\calE_{\rm fl,fin}(t,T)$ (short-dashed line) and total
energy (solid line) for parameter set 1. \\
{\bf Fig. 1d:} The pressure $p(t)$ for parameter set 1. \\
{\bf Fig. 1e:} Particle number $n(t)-n(0)$ for parameter set 1. \\
{\bf Fig. 2a:} $\phi(t)$ for parameter set 2. \\
{\bf Fig. 2b:} The potential $\calv(t)$ for parameter set 2. \\
{\bf Fig. 3a:} $\phi(t)$ for parameter set 3. \\
{\bf Fig. 3b:} The potential $\calv(t)$ for parameter set 3. \\
{\bf Fig. 4a:} $\phi(t)$ for parameter set 4. \\
{\bf Fig. 4b:} The potential $\calv(t)$ for parameter set 4. \\
{\bf Fig. 4c:} Particle number $n(t)-n(0)$ for parameter set 4.  

\vspace*{20mm}
\parbox{15cm}{
\begin{center}
\begin{tabular}{|c|c|c|c|c|c|c|}
\hline
set \# &$\lambda$&$T$&$m_0$&$\phi(0)$&$\calm^2(\infty)$&r.h.s.\\
\hline
1&1&0&3&2.815&5.0&4.98\\
2&5&0&3&1.235&4.95&4.90\\
3&1&3&3&2.76&5.15&5.39\\
4&1&10&3&1.25&8.24&9.45\\
\hline
\end{tabular} \end{center}}
\begin{center}
{\bf Table 1}
\end{center}

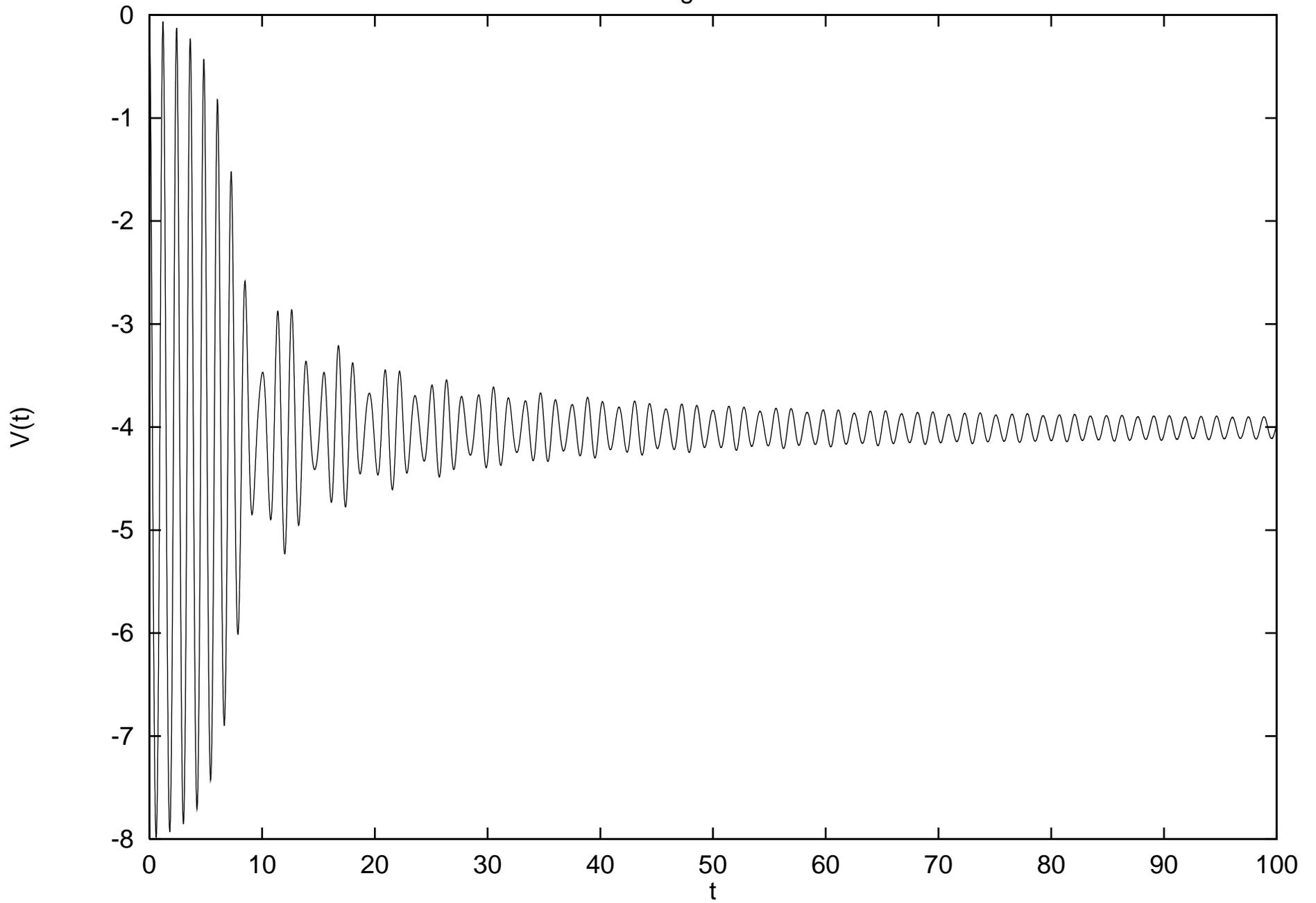
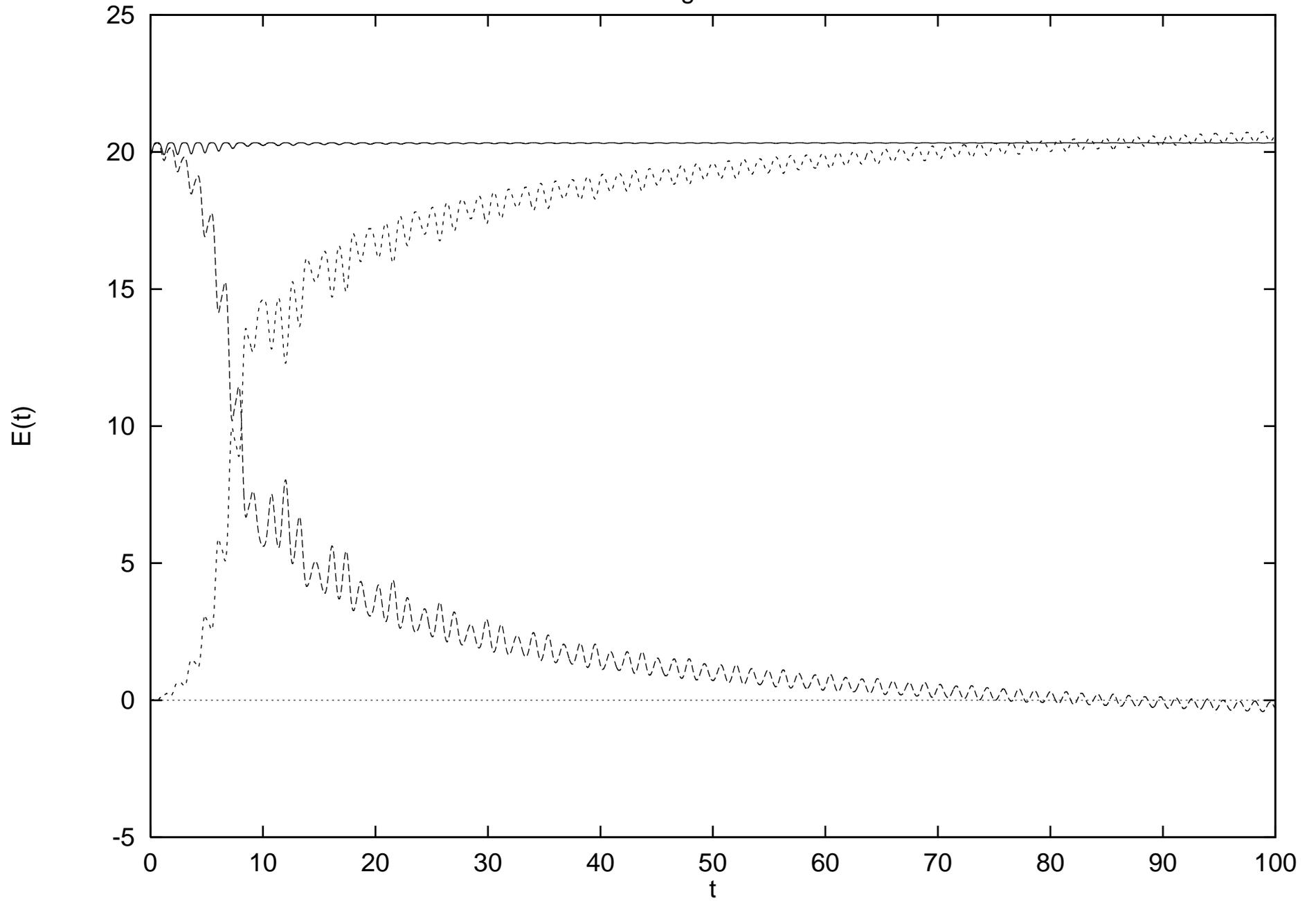
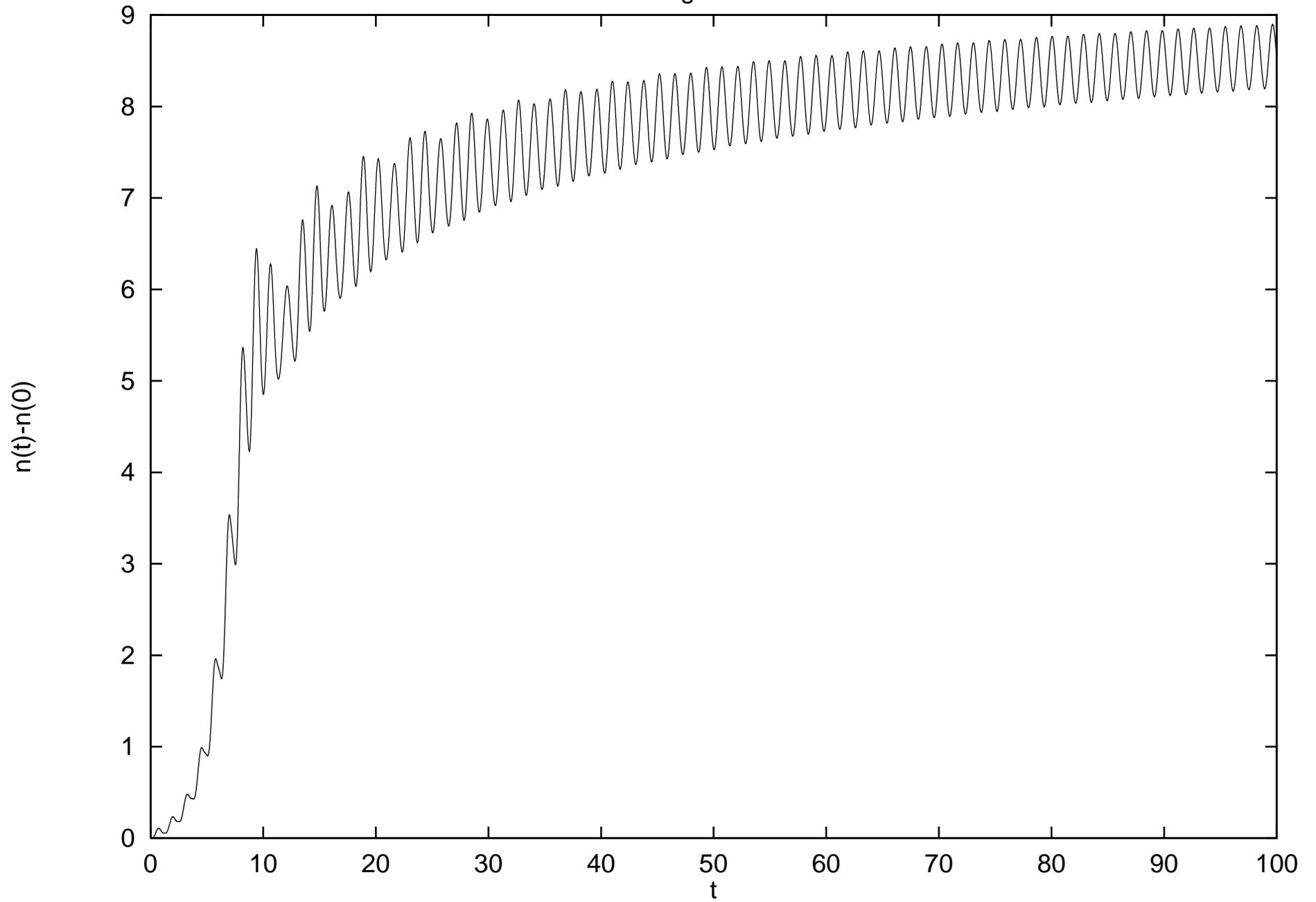
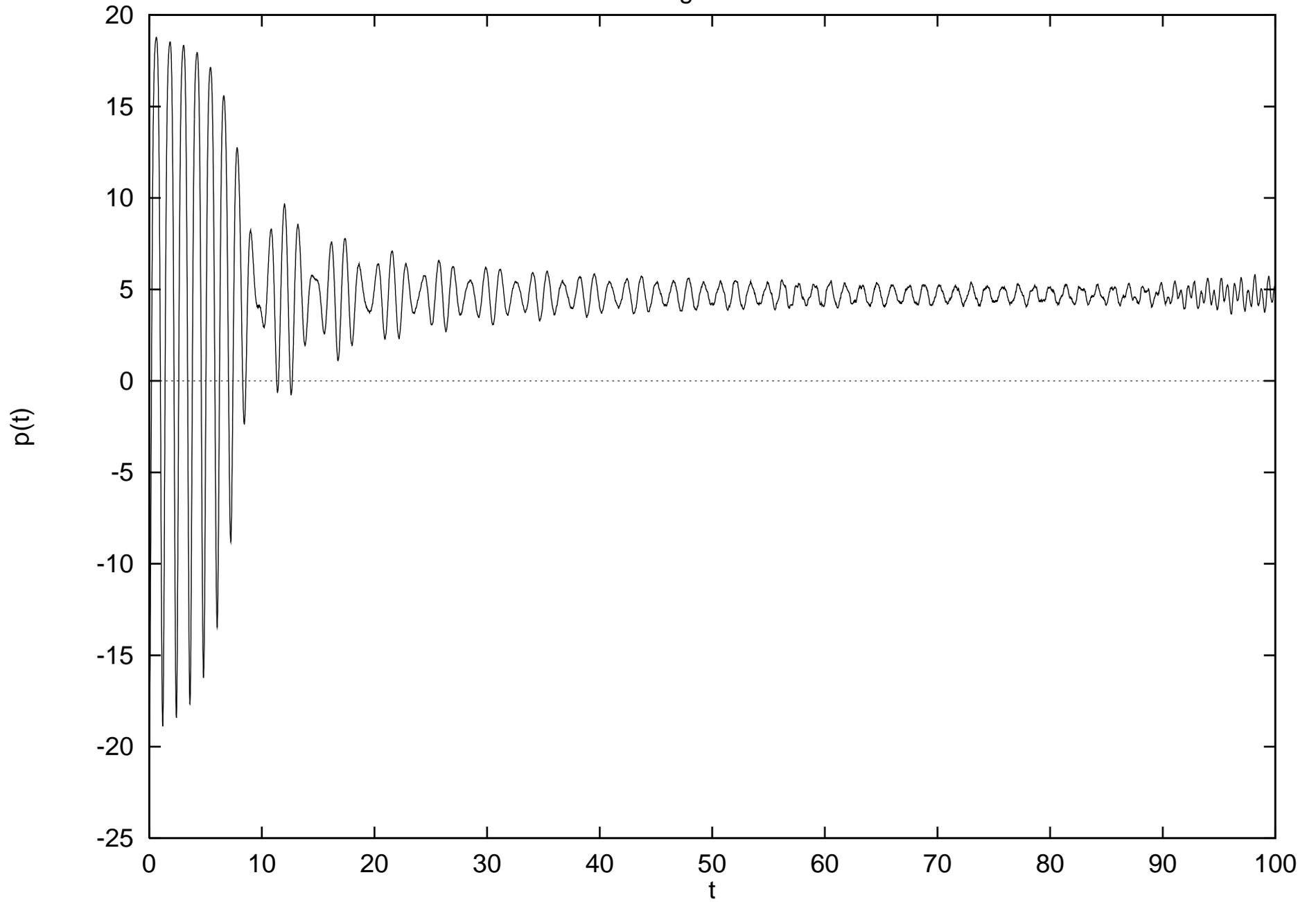
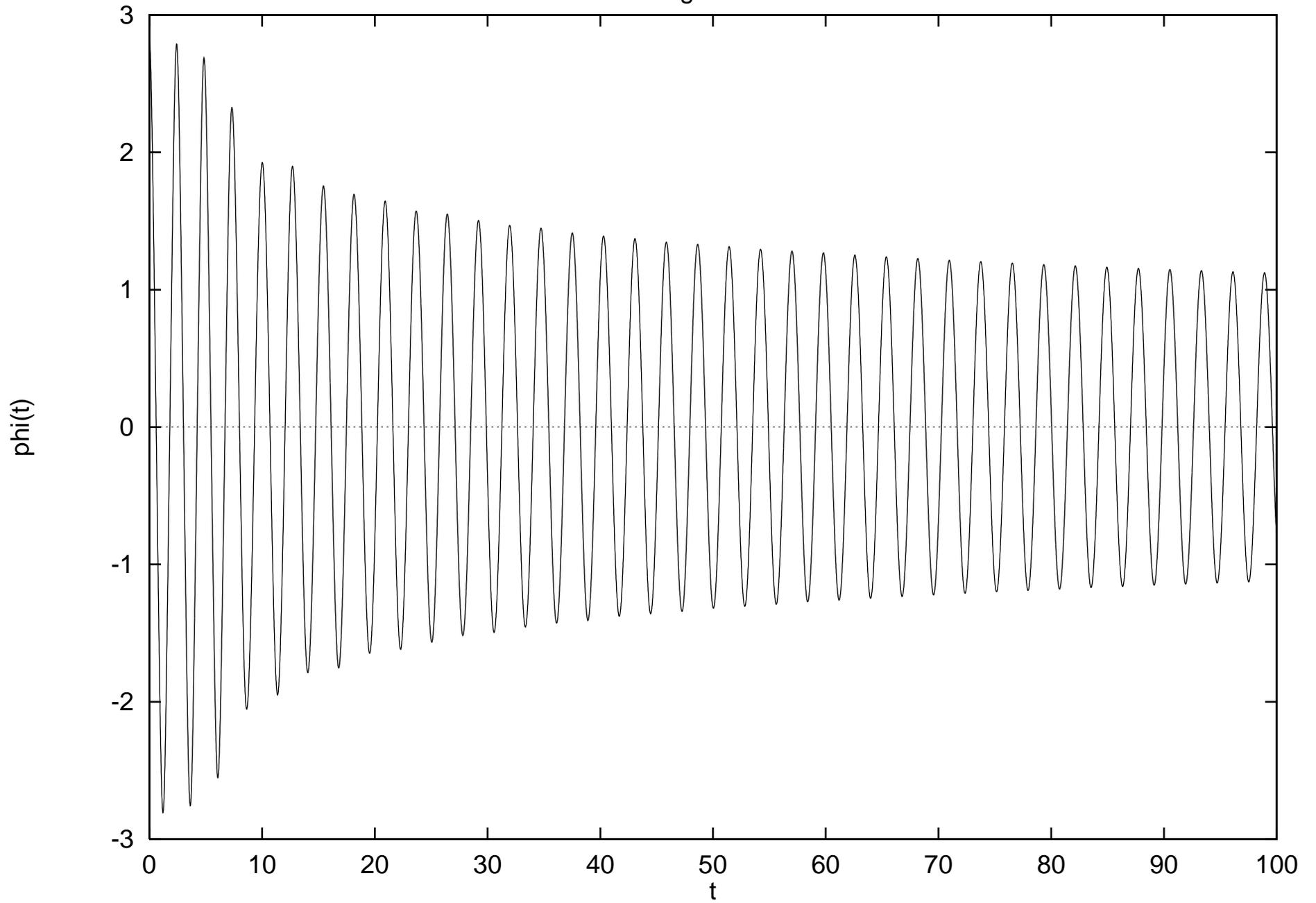
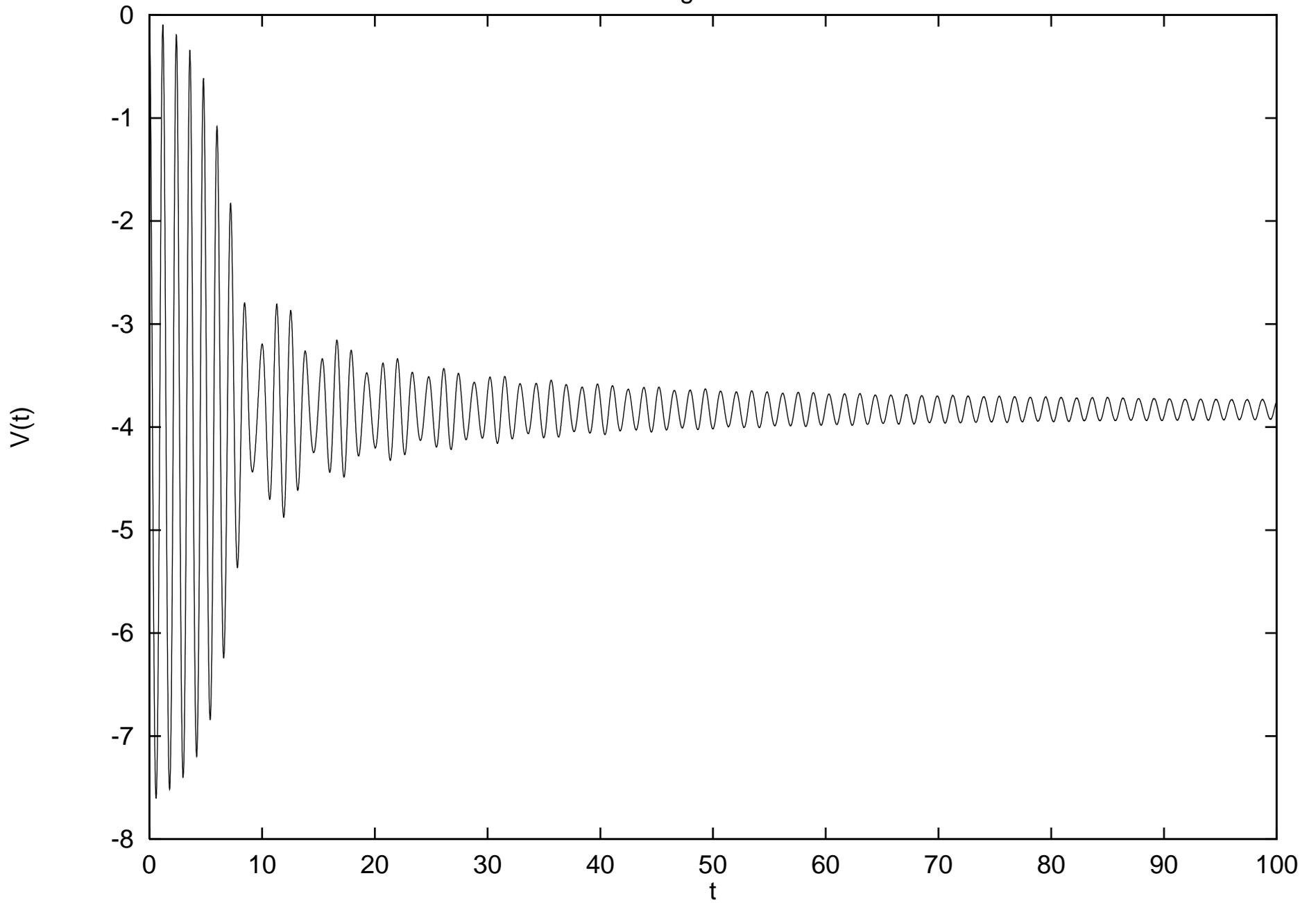
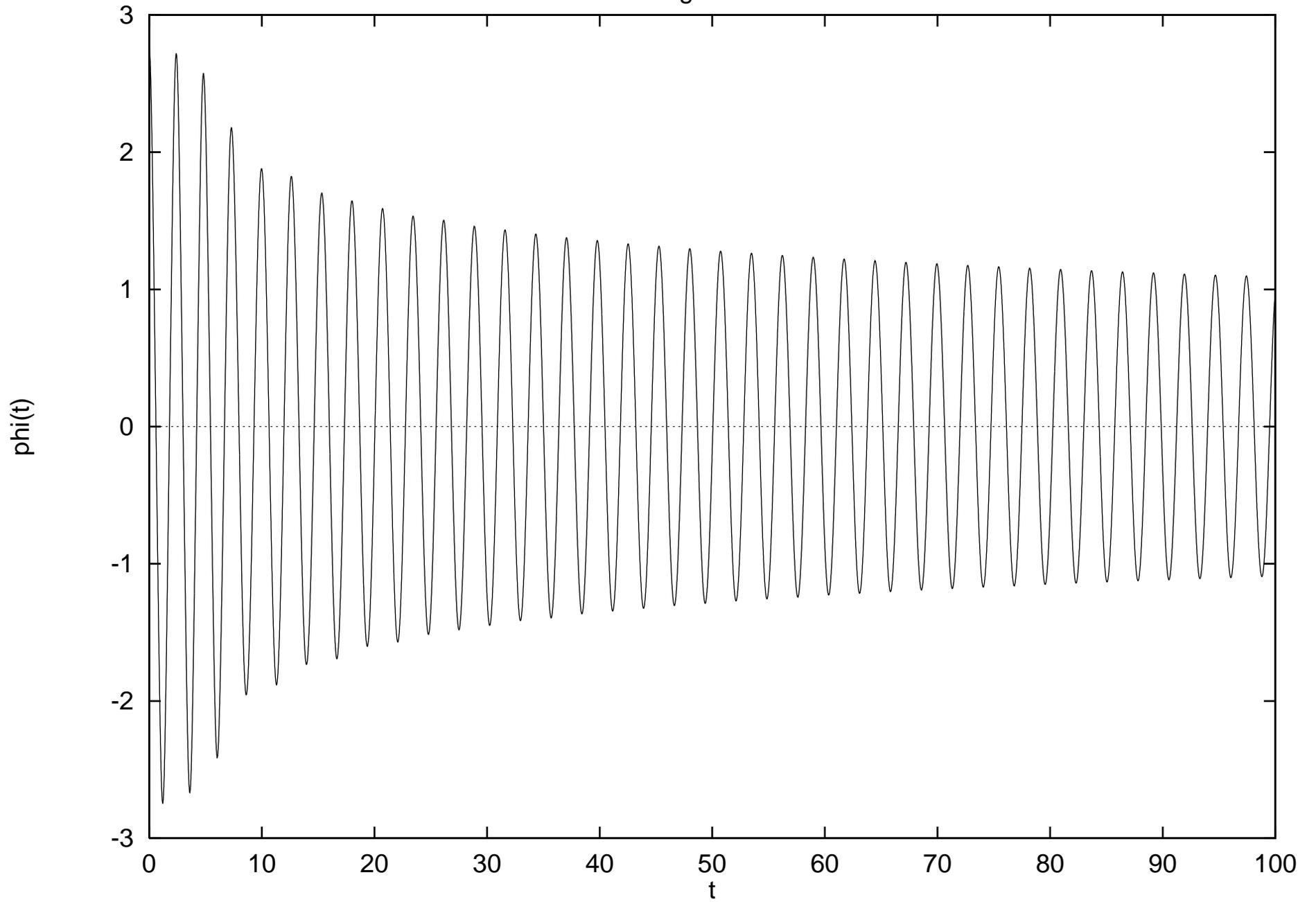
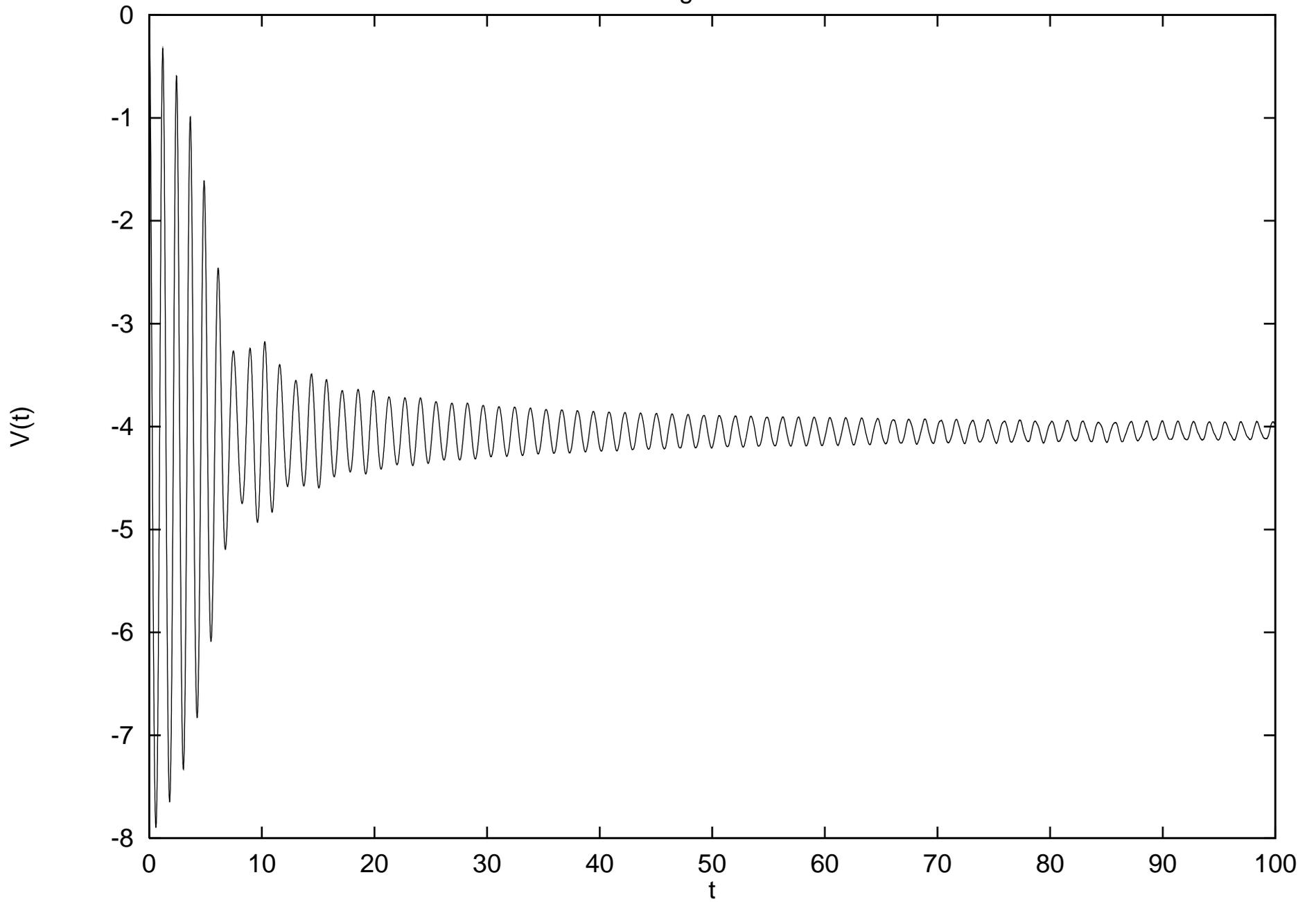
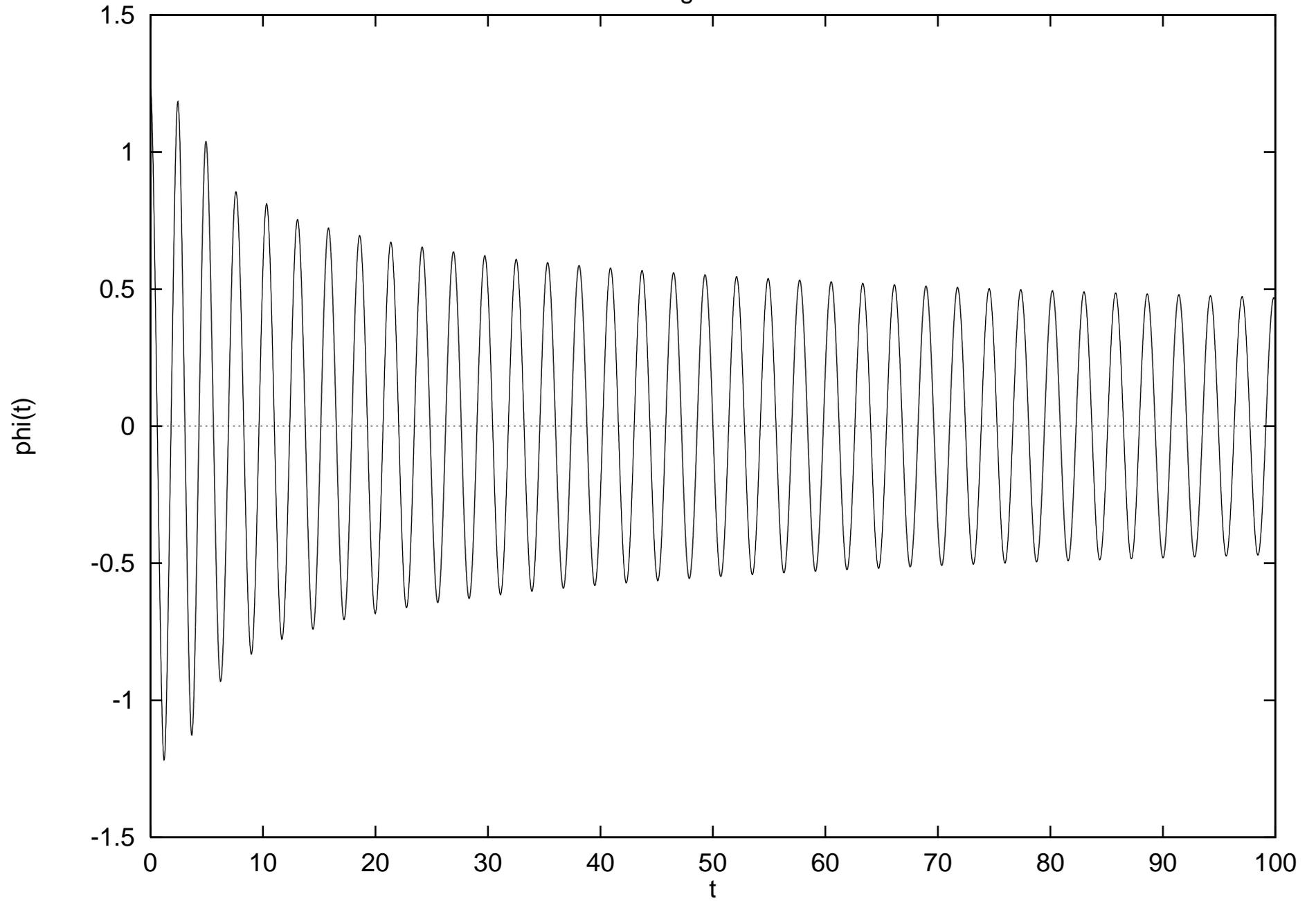
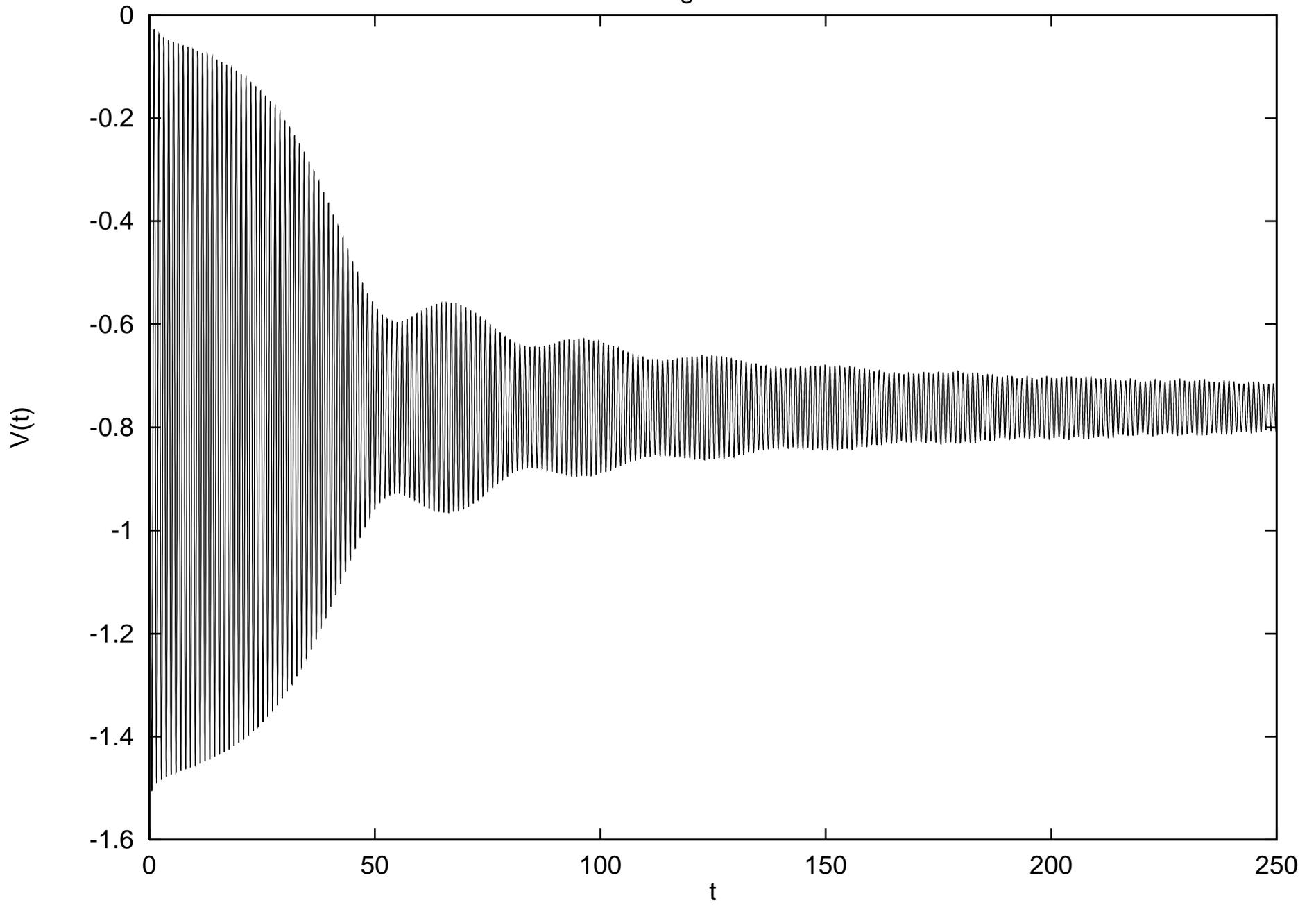
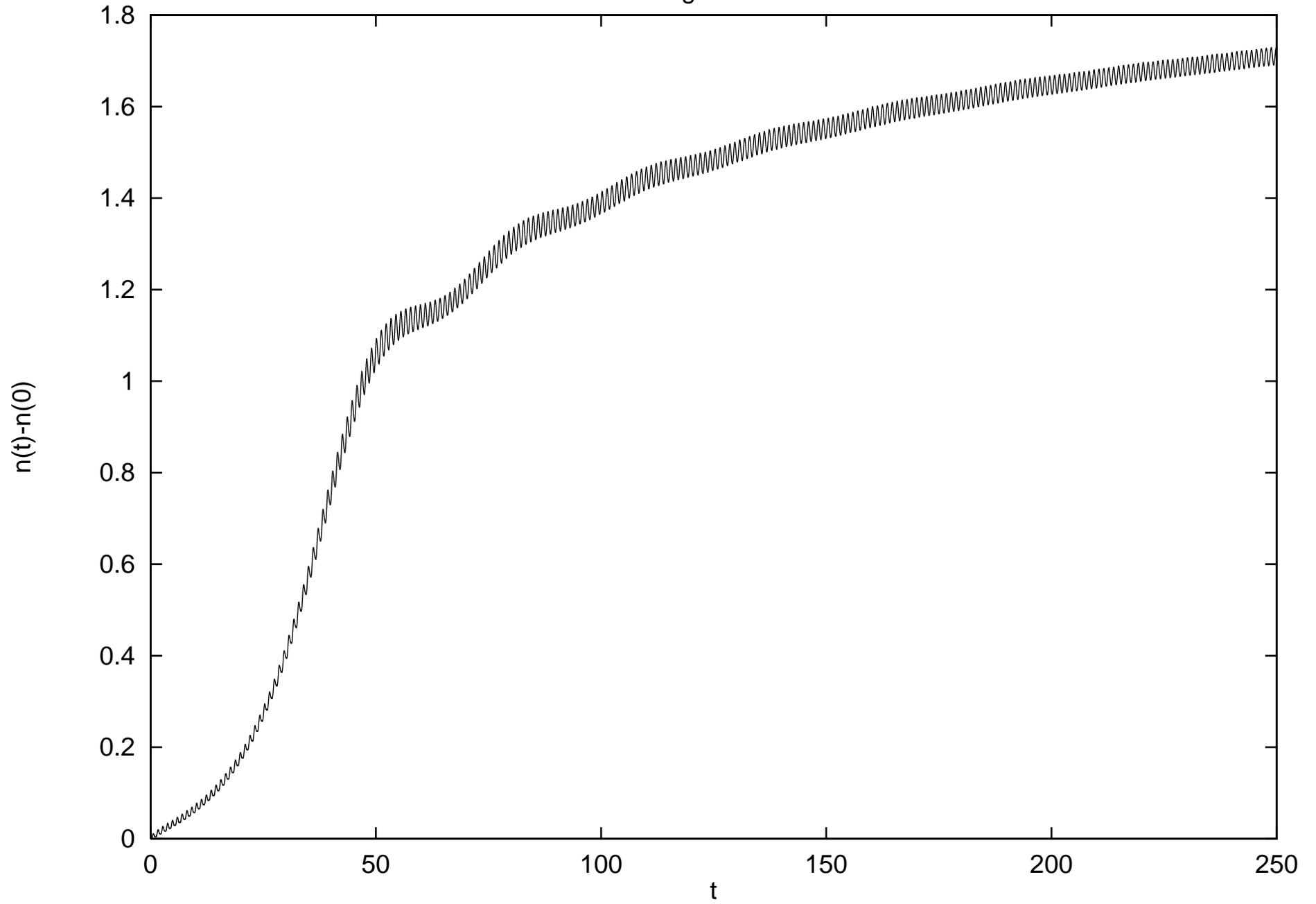
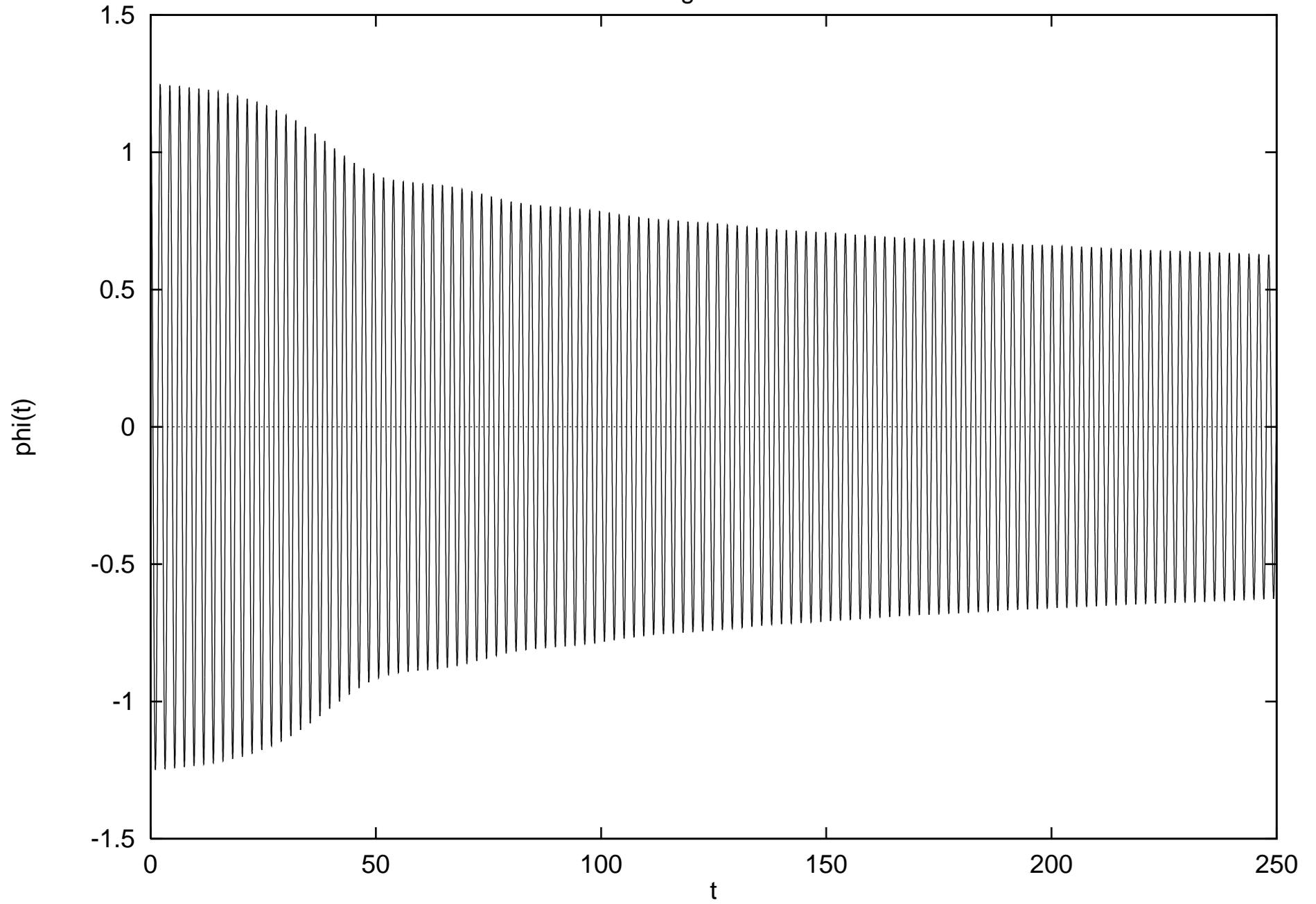
\end{document}